\documentclass{pssbmac}

\usepackage[english]{babel} 
\usepackage{bm}
\usepackage[utf8]{inputenc} 

\usepackage{listings}
\usepackage{xcolor}
\usepackage[colorlinks,linkcolor=blue]{hyperref}

\usepackage[T1]{fontenc}
\usepackage{float}
\usepackage{graphics}
\usepackage{graphicx}
\usepackage{epsfig}
\usepackage{indentfirst}
\usepackage{amsmath, amsfonts, amssymb, amsthm, mathtools}
\usepackage{url}
\usepackage{csquotes}

\usepackage[backend=biber, style=ieee, maxnames=50]{biblatex}
\DeclareTextFontCommand{\emph}{\boldmath\bfseries}
\DefineBibliographyStrings{english}{mathesis = {Master dissertation}}

\begin{document}


\title{Implementation of the Feedforward Multichannel Virtual Sensing Active Noise Control (MVANC) by Using MATLAB}

\author{
    {\large Boxiang Wang}\thanks{BOXIANG001@e.ntu.edu.sg} 
}
\criartitulo


\begin{abstract}
The multichannel virtual sensing active noise control (MVANC) methodology is an advanced approach that may provide a wide area of silence at specific virtual positions that are distant from the physical error microphones. Currently, there is a scarcity of open-source programs available for the MVANC algorithm. This work presents a MATLAB code for the MVANC approach, utilizing the multichannel filtered-x least mean square (MCFxLMS) algorithm. The code is designed to be applicable to systems with any number of channels. The code can be found on \href{https://github.com/ShiDongyuan/Multichannel_Virtual_Sensing_ANC}{GitHub}. 

\end{abstract}

\section{Introduction}
Active noise control, often known as ANC, is a cutting-edge method that eliminates unwanted noise by providing regulated anti-noise that effectively cancels out the noise that is present in the environment. In the field of wave physics, this phenomenon, which is referred to as destructive interference, is an advanced application of the theory of superposition~\cite{shi2023active1,lam2021ten}. With the help of the adaptive algorithms~\cite{guo2024survey,ji2023computation,lai2023mov,ji2023practical,shi2021optimal,shi2021comb}, the use of active noise cancellation (ANC) is widespread in a variety of applications that are sensitive to noise problems. Some instances of these applications are windows~\cite{lam2023anti,shi2019practical,lam2020active, hasegawa2018window,lam2018active,shi2023computation2,lai2023robust,shi2016open,shi2021block,lam2020active1,shi2017algorithmsC,shi2017understanding,luo2024real,shen2024principle}, vehicles, and headphones~\cite{shen2022hybrid,shen2022multi,shi2022selective,shen2021alternative,shen2023implementations,shen2021wireless,shen2022adaptive,shi2023transferable}. ANC is also efficient at minimizing low-frequency noise~\cite{shi2019two,gan2023practical}.
However, it is essential to note that the "quiet zone" is only localized to the monitoring "error sensor" in the majority of ANC applications. This is because the monitoring "error sensor" supplies a posteriori information to the adaptive algorithm. Therefore, the typical approach to active noise control (ANC), in which the quiet zone is dependent on the positioning of the physical error sensor, appears to be unworkable in situations where physical placement is inconvenient, such as at the eardrums~\cite{miyazaki2015head,pawelczyk2008analog,pawelczyk2003noise,rafaely1999broadband,deng2018active} or domestic living areas~\cite{hasegawa2018window}. Currently, numerous forms of virtual sensing ANC approaches have been presented in order to ease the restriction on error sensor placement. This problem cannot be solved without these techniques~\cite{kidner2006feasibility}. The silent zone can be efficiently generated in these ways at the desired 'virtual' locations, which are located away from the actual error sensors~\cite{qiu2018review}.
Specifically, we focus on a virtual sensing ANC algorithm that uses an auxiliary filter to achieve optimal control on the virtual position which overcomes the spatial correlation and causality constraint between error microphones~\cite{shi2020feedforward}. 
This technique requires a preliminary training stage to produce an auxiliary filter that indirectly records the relationship between the physical sensor and the required virtual location. This stage is necessary to obtain the filter. As a consequence, the pre-trained filter assists the ANC system in obtaining sufficient management of the disturbance at the position of the virtual error sensor.
To obtain a larger quiet zone at the desired location, this study realizes the feedforward multichannel virtual sensing ANC (MVANC) technique employing the widely-used simulation tool, MATLAB, and makes its code accessible to the public.

A comprehensive explanation of a MATLAB program that simulates a feedforward multichannel virtual sensing active noise control (MVANC) system by utilizing the multichannel filtered-x least mean square (MCFxLMS)~\cite{luo2022implementation,shi2023multichannel,shi2020active,shi2024behindN} method is provided in this paper. 

\section{Theoretical Introduction on Feedforward Multichannel Virtual Sensing ANC Technique}

\begin{figure}[H]
\centering
\includegraphics[width=.6\textwidth]{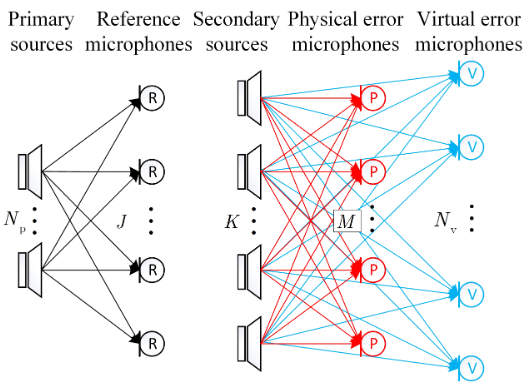}
\caption{ {\small Simplified schematic of the MVANC algorithm.}}
\label{figure 1}
\end{figure}

Figure~\ref{figure 1} depicts the implementation of a generic multichannel ANC system, which includes reference microphones with a value of $J$, secondary sources with a value of $K$, and physical error microphones with a value of $M$. A further point to consider is that there are $N_p$ independent primary sources that are positioned ahead of the reference microphones. When applied to this multichannel system, the MVANC approach is utilized to cancel out the disturbances that occur at the $N_v$ virtual microphone positions in the distant field. It is assumed that the number of reference microphones is more than the number of primary sources ($N_p < J$). This assumption is made for the sake of brevity and to avoid losing generality.

The MVANC technique is comprised of two stages: the first stage is the tuning stage, and the second stage is the control stage. During the tuning stage, the output signal vector is determined by the following equation, as depicted to the right in Figure 2: 
\begin{equation}
{\bm{y}}(n) = {{\bm{w}}^{\bm{T}}}(n){\bm{x}}(n),
\label{Eq:1}
\end{equation}
where 
$${\bm{y}}(n) = {[{y_1}(n),{y_2}(n),...,{y_k}(n),...,{y_K}(n)]^T}.$$
The expression ${y_k}(n)$ represents the kth output signal at a time sample of $n$, while the expression ${( \cdot )^T}$ represents the matrix transpose operator. The stacked reference vector is stated to be
\begin{equation}
{\bm{x}}(n) = {[{\bm{x}}_1^T(n),{\bm{x}}_2^T(n),...,{\bm{x}}_j^T(n),...,{\bm{x}}_J^T(n)]^T},
\label{Eq:2}
\end{equation}
in which ${{\bm{x}}_j}(n)$ denotes the $j$th reference vector, which owns ${N_x}$ elements and is expressed as  

\begin{equation}
\bm{x}_j(n) = \left[ x_j(n), x_j(n - 1), \ldots, x_j(n - N_x + 1) \right]^\text{T}.
\label{Eq:3}
\end{equation}

The control filter matrix ${\bm{w}}(n)$ in Eq.~(\ref{Eq:1}) is given by
\begin{equation}
\bm{w}(n) = 
\begin{bmatrix}
\bm{w}_{11}(n) & \bm{w}_{21}(n) & \cdots & \bm{w}_{K1}(n) \\
\bm{w}_{12}(n) & \bm{w}_{22}(n) & \cdots & \bm{w}_{K2}(n) \\
\vdots & \vdots & \ddots & \vdots \\
\bm{w}_{1J}(n) & \bm{w}_{2J}(n) & \cdots & \bm{w}_{KJ}(n)
\end{bmatrix},
\label{Eq:4}
\end{equation}
where the variable ${{\bm{w}}_{kj}}(n)$ denotes the weight vector of the control filter, which determines the influence of the $j$th input reference signal on the $k$th output control signal: 

\begin{equation}
\bm{w}_{kj}(n) = \left[ w_{kj,0}(n), w_{kj,1}(n), \ldots, w_{kj,N_x-1}(n) \right]^\text{T}.
\label{Eq:5}
\end{equation}

According to the traditional MCFxLMS technique, the new control filter that extends from the $j$th input to the $k$th output is obtained from 
\begin{equation}
{{\bm{w}}_{kj}}(n + 1) = {{\bm{w}}_{kj}}(n) - {\mu _1}\sum\limits_{i = 1}^{{N_v}} {{{{\bm{x'}}}_{v,jki}}(n){e_{v,i}}(n)},
\label{Eq:6}
\end{equation}
where the symbol $\mu_1$ is used to indicate the stepsize of the MCFxLMS algorithm, while the symbol ${{e_{v,i}}(n)}$ is used to represent the error signal at the $i$th virtual microphone. The filtered reference ${{{x'}_{v,jki}}(n)}$ is a convolution that is performed between the $j$th reference signal ${x_j}(n)$ and the virtual secondary path estimate ${{\hat g}_{v,ik}}(n)$, which includes the path from the $k$th secondary source to the $i$th virtual microphone. In the tuning step, it is possible to observe that the final control filter converges to the same optimal control filter that was achieved using the standard MCFxLMS. This knowledge may be derived from the equation~\eqref{Eq:6}.


Following this procedure, the optimal control filters can be obtained once the training converges. Then, the auxiliary filters are trained using the least mean square (LMS) algorithm as shown in Fig.~\ref{figure 3}. The $m$th auxiliary filter vector can be obtained from 
\begin{equation}
\bm{h}_m(n + 1) = \bm{h}_m(n) + \mu_2 e_{h,m}(n) \bm{{\bar x}}(n),
\label{Eq:7}
\end{equation}

\begin{figure}[H]
\centering
\includegraphics[width=.7\textwidth]{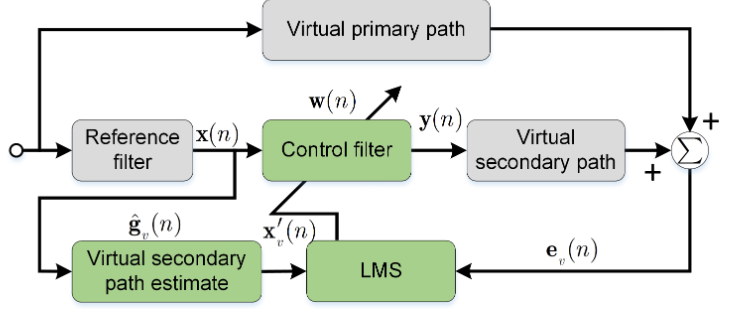}
\caption{ {\small Training the control filter during the tuning phase of the MVANC process.}}
\label{figure 2}
\end{figure}
\noindent
where $\mu_2$ denotes the stepsize of the LMS algorithm, and 
\begin{equation}
\bm{h}_m(n) = \left[
\bm{h}_{m1}^T(n), \,
\bm{h}_{m2}^T(n), \,
\ldots, \,
\bm{h}_{mj}^T(n), \,
\ldots, \,
\bm{h}_{mJ}^T(n)
\right]^T.
\label{Eq:8}
\end{equation}

The expression $\bm{h}_{mj}(n)$ denotes the auxiliary filter from the $j$th reference microphone to the $m$th physical microphone with a length of ${N_h}$ taps. Therefore,
\begin{equation}
\bm{h}_{mj}(n) = \left[
h_{mj,0}(n), \,
h_{mj,1}(n), \,
\ldots, \,
h_{mj,N_h-1}(n)
\right]^\text{T}.
\label{Eq:9}
\end{equation}

The stacked reference vector is given by 
\begin{equation}
\bar{\bm{x}}(n) = \left[
\bar{\bm{x}}_1^\text{T}(n), \,
\bar{\bm{x}}_2^\text{T}(n), \,
\ldots, \,
\bar{\bm{x}}_J^\text{T}(n)
\right]^\text{T},
\label{Eq:10}
\end{equation}
where $\bar{\bm{x}}_j(n)$ represents the $j$th reference vector with ${N_h}$ elements and is expressed as
\begin{equation}
\bar{\bm{x}}_j(n) = \left[
x_j(n), \,
x_j(n - 1), \,
\ldots, \,
x_j(n - N_h + 1)
\right]^\text{T}.
\label{Eq:11}
\end{equation}

The error signal ${e_{h,m}}(n)$ is stated to be 
\begin{equation}
e_{h,m}(n) = e_{p,m}(n) - \bm{h}_m^\text{T}(n)\bar{\bm{x}}(n),
\label{Eq:12}
\end{equation}

\begin{figure}[H]
\centering
\includegraphics[width=.7\textwidth]{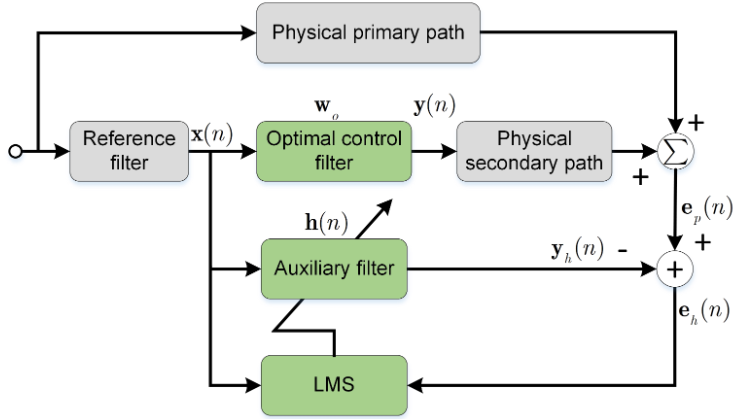}
\caption{ {\small Training the auxiliary filter during the tuning phase of the MVANC technique.}}
\label{figure 3}
\end{figure}

\begin{figure}[H]
\centering
\includegraphics[width=.7\textwidth]{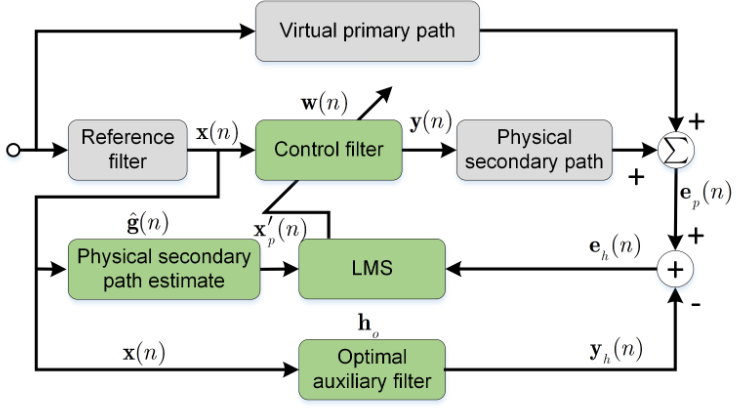}
\caption{ {\small Training the control filter during the control phase of the MVANC technique.}}
\label{figure 4}
\end{figure}
\noindent
where $e_{p,m}(n)$ denotes the error signal at the $m$th physical microphone.

During the control step, the physical microphones are retained while the virtual microphones are eliminated. As depicted in Figure~\ref{figure 4}, the new control filters are obtained by utilizing the MCxLMS algorithm.

\begin{equation}
{{\bm{w}}_{kj}}(n + 1) = {{\bm{w}}_{kj}}(n) - {\mu _3}\sum\limits_{m = 1}^M {{{{\bm{x'}}}_{p,jkm}}(n){e_{h,m}}(n)} ,
\label{Eq:13}
\end{equation}
where 
the symbol $\mu_3$ represents the stepsize of the MCFxLMS algorithm. The variable ${{x'}_{p,jkm}}(n)$ represents the product of the convolution between the secondary path estimate ${{\hat g}_{mk}}(n)$, which is derived from the $k$th secondary source to the $m$th physical microphone, and the $j$th reference signal represented by ${x_j}(n)$.
The error signal $e_{h,m}$ is given by 

\begin{equation}
e_{h,m}(n) = e_{p,m}(n) - \bm{h}_{0,m}^T \bar{\bm{x}}(n),
\label{Eq:14}
\end{equation}
where $\bm{h}_{o,m}$ represents the $m$th optimal auxiliary filter obtained from Eq.~(\ref{Eq:7}).

Following the tuning stage and control stage described earlier, the feedforward MVANC system that has been proposed is capable of achieving noise cancellation at $N$ virtual places inside the system.

\section{Code Explanation}
The section provides the explanation of the program of a feedforward MVANC in Matlab. This program includes the following five MATLAB files: \textcolor{blue}{$CreatReferenceSignal.m$} generates the filtered reference signals and disturbances, \textcolor{blue}{$MultichannelFxLMS.m$}, \textcolor{blue}{$AuxiliaryLMS.m$} and \textcolor{blue}{$ContrFxLMS.m$} are the programs of the MVANC technique, while \textcolor{red}{$VirtualSensing\_test.m$} is the main program used to evaluate the programs of the MVANC system.  

\subsection{Function: CreatReferenceSignal}
The MVANC technique utilizes the FxLMS algorithm to adaptively update the control filter coefficients. Therefore, \textcolor{blue}{$CreatReferenceSignal.m$} generates the reference signals filtered by the physical secondary path and the virtual secondary path respectively. In addition, it produces disturbances at the location of the physical microphone and virtual microphone respectively. These signals will later be used in other functions.

\begin{lstlisting}[language=Matlab]
function [Dv,Dp,Fx_v,Fx_p] = CreatReferenceSignal(Pv,Pp,Sv,Sp,PriNoise,N,L,K,M,J)
%% -----------------------------------------------------------
% Inputs:
% Pv is the virtual primary path. 
% Pp is the physical primary path. 
% Sv is the virtual secondary path.
% Sp is the physical secondary path.
% PriNoise is the primary noise. 
% N is the number of simulation cycles.
% L is the length of control filter.
% K is the number of secondary sources.
% M is the number of virtual microphones.
% J is the number of physical microphones.
% Outputs:
% Dv is the disturbance at virtual microphone.
% Dp is the disturbance at physical microphone.
% Fx_v is the reference signal filtered by virtual secondary path.
% Fx_p is the reference signal filtered by physical secondary path.
%% -----------------------------------------------------------
Dp = zeros(N,J);
for i=1:J
    Dp(:,i) = filter(Pp(:,i),1,PriNoise);
end
Dv = zeros(N,M);
for i=1:M
    Dv(:,i) = filter(Pv(:,i),1,PriNoise);
end
Fx_p = zeros(N+L-1,J,K);
for i = 1:K 
    for j = 1:J
        Fx_p(:,j,i) = [zeros(L-1,1);filter(Sp(:,j,i),1,PriNoise)];
    end
end
Fx_v = zeros(N+L-1,M,K); 
for i = 1:K 
    for j = 1:M
        Fx_v(:,j,i) = [zeros(L-1,1);filter(Sv(:,j,i),1,PriNoise)]; 
    end
end
end
\end{lstlisting}

In this code snippet, $Dv$ is used to store the disturbances at virtual microphones and has a dimension of $N$ by $M$, while $Dp$ is used to store the disturbances at physical microphones and has a dimension of $N$ by $J$. In addition, $Fx\_v$ is used to store the reference signals filtered by virtual secondary paths and has a dimension of $N$ by $M$ by $K$, while $Fx\_p$ is used to store the reference signals filtered by virtual secondary paths and has a dimension of $N$ by $J$ by $K$. $N$, $M$, $J$ and $K$ denote the simulation cycle number, the number of virtual error microphones, the number of physical error microphones and the number of secondary sources respectively.

\subsection{Function: MultichannelFxLMS}
In the tuning stage of the MVANC technique, the optimal control filters are first trained by \textcolor{blue}{$MultichannelFxLMS.m$} using the FxLMS algorithm.
\begin{lstlisting}[language=Matlab]
function [W,Er]= MultichannelFxLMS(L,K,M,N,Fx_v,Dv,StepSize)
%% -----------------------------------------------------------
% Inputs:
% L is the length of control filter.
% K is the number of secondary sources.
% M is the number of virtual microphones. 
% N is the number of simulation cycles.
% Fx_v is the reference signal filtered by virtual secondary path. 
% Dv is the disturbance at virtual microphone.
% StepSize is the stepsize of FxLMS algorithm.
% Outputs:
% W is the control filter matrix.
% Er is the error signal at virtual microphone.
%% -----------------------------------------------------------
W  = zeros(K*L,1); 
FX = zeros(M,K*L); 
Er = zeros(M,N); 
for i = 1:N
    for j = 1:M
        for kk= 1:K
            FX(j,(kk-1)*L+1:kk*L) = Fx_v(i+L-1:-1:i,j,kk)';
        end
    end
    Ev = Dv(i,:)'- FX*W;
    SX = StepSize*Ev'*FX;
    W  = W + SX';
    Er(:,i) = Ev;
end
end
\end{lstlisting}

In this code snippet, $W$ is used to store the optimal control filters trained using the FxLMS algorithm and has a dimension of $K$ by $L$, where $L$ denotes the length of the control filter. $Er$ is used to store the error signals at virtual microphones and has a dimension of $M$ by $N$.
\newpage

\subsection{Function: AuxiliaryLMS}
After the control filter converges, the optimal control filters are then used by \textcolor{blue}{$AuxiliaryLMS.m$} to train the auxiliary filters using the LMS algorithm.
\begin{lstlisting}[language=Matlab]
function [H,Er]=AuxiliaryLMS(L,K,J,N,Fx_p,Dp,PriNoise,W,StepSize)
%% -----------------------------------------------------------
% Inputs:
% L is the length of control filter.
% K is the number of secondary source.
% J is the number of physical microphones. 
% N is the number of simulation cycle.
% Fx_p is the reference signal filtered by physical secondary path.
% Dp is the disturbance at physical microphone.
% W is the optimal control filter matrix.
% StepSize is the stepsize of LMS algorithm.
% Outputs:
% H is the auxiliary filter matrix.
% Er is the error signal of LMS algorithm.
%% -----------------------------------------------------------
H  = zeros(J*L,1); 
SH = zeros(J*L,1); 
FX = zeros(J,K*L); 
YH = zeros(J,1); % The output signal of the auxiliary fitler.
Er = zeros(J,N)  ;
X  = [zeros(L-1,1); PriNoise];
for i = 1:N
    for j = 1:J
        for kk = 1:K 
            FX(j,(kk-1)*L+1:kk*L) = Fx_p(i+L-1:-1:i,j,kk)';
        end
        YH(j) =  X(i+L-1:-1:i)'*H((j-1)*L+1:j*L);
    end
    Ep = Dp(i,:)'-FX*W; % The error signal at physical microphone.
    Eh = Ep - YH;
    for j = 1:J
        SH((j-1)*L+1:j*L) = StepSize*Eh(j)*X(i+L-1:-1:i);
    end
    H = H + SH;
    Er(:,i) = Eh;
end
end
\end{lstlisting}

In this code snippet, $H$ is used to store the optimal auxiliary filters trained using the LMS algorithm and has a dimension of $J$ by $L$. $Er$ is used to store the error signals and has a dimension of $J$ by $N$.

\newpage

\subsection{Function: ContrFxLMS}
In the control stage of the MVANC technique, the optimal auxiliary filers are used by \textcolor{blue}{$ContrFxLMS.m$} to train the new control filters using the FxLMS algorithm.
\begin{lstlisting}[language=Matlab]
function [WC,ErPhysic,ErVirt] = ContrFxLMS(L,K,M,J,N,Fx_p,Fx_v,Dp,Dv,PriNoise,H,StepSize) 
%% ------------------------------------------------------------------------
% Inputs:
% L is the length of control filter.
% K is the number of secondary source.
% M is the number of virtual microphones. 
% J is the number of physical microphones. 
% N is the number of simulation cycle.
% Fx_p is the reference signal filtered by physical secondary path.
% Fx_v is the reference signal filtered by virtual secondary path. 
% Dp is the disturbance at physical microphone.
% Dv is the disturbance at virtual microphone.
% H is the optimal auxiliary filter matrix.
% StepSize is the stepsize of FxLMS algorithm.
% Outputs:
% WC is the new control filter matrix.
% ErPhysic is the error signal at physical microphone.
% ErVirt is the error signal at virtual microphone.
%% ------------------------------------------------------------------------
FX = zeros(J,K*L);
FV = zeros(M,K*L);
WC = zeros(K*L,1); 
ErPhysic = zeros(J,N); 
ErVirt   = zeros(M,N); 
XH       = zeros(N,J);
for i = 1:J 
    XH(:,i) = filter(H((i-1)*L+1:i*L),1,PriNoise);
end
for i=1:N
    for j = 1:J
        for kk = 1:K 
            FX(j,(kk-1)*L+1:kk*L) = Fx_p(i+L-1:-1:i,j,kk)';
        end
    end
    for j = 1:M
        for kk = 1:K 
            FV(j,(kk-1)*L+1:kk*L) = Fx_v(i+L-1:-1:i,j,kk)';
        end
    end
    Ep      = Dp(i,:)' - FX*WC; 
    ErVirt(:,i) = Dv(i,:)' - FV*WC; 
    Eh      = Ep-XH(i,:)';
    WC      = WC + (StepSize*Eh'*FX)';
    ErPhysic(:,i) = Ep;
end
end
\end{lstlisting}

In this code snippet, $WC$ is used to store the new control filters trained using the FxLMS algorithm and has a dimension of $K$ by $L$. $ErPhysic$ is used to store the error signals at physical microphones which have a dimension of $J$ by $N$, while $ErVirt$ is used to store the error signals at virtual microphones which have a dimension of $M$ by $N$.

\section{Testing Code: Four Channel Virtual Sensing Active Nose Control System}
The \textcolor{red}{$VirtualSensing\_test.m$} carries out a simulation on a 1x4x4 MVANC system, where there is $1$ reference microphone, $4$ secondary sources, $4$ physical microphones and $4$ virtual microphones. In the simulation, \textcolor{blue}{$CreatReferenceSignal.m$}, \textcolor{blue}{$MultichannelFxLMS.m$}, \textcolor{blue}{$AuxiliaryLMS.m$} and \textcolor{blue}{$ContrFxLMS.m$} are used to achieve noise control at desired virtual locations.

\subsection{System Configuration}
The following commands define basic system configuration parameters. Specifically, the sampling frequency of the system is set to 16 kHz and the length of control filters is set to 512 taps.
\begin{lstlisting}[language=Matlab]
Fs         = 16000       ; % System sampling freqnecy (Hz).
T          = 50          ; % The duration of the simutation (Second).
t          = 0:1/Fs:T    ; % The cycle of simulation. 
N          = length(t)   ; % The number of cycle of simulation. 
L          = 512         ; % The lenght of the control filter. 
K          = 4           ; % The number of the seconeary sources. 
J          = 4           ; % The number of the physical microphones.
M          = 4           ; % The number of the virtual microphones.
\end{lstlisting}

\subsection{Create Primary Noise for Tuning and Control}
The following commands generate a broadband noise of 800-2500 Hz as reference signals in the tuning stage and a broadband noise of 800-1800 Hz as reference signals in the control stage by filtering white noise through different bandpass filters.

\begin{lstlisting}[language=Matlab]
% Creating primary noise for tuning.
Fs1_tuning  = 800;
Fs2_tuning  = 2500;
ws1_tuning  = 2*(Fs1_tuning/Fs);
ws2_tuning  = 2*(Fs2_tuning/Fs);
BPF_tuning = fir1(512,[ws1_tuning,ws2_tuning]);
PriNoise_tuning = filter(BPF_tuning,1,2*randn(N,1));
% Creating primary noise for control.
Fs1_control  = 800;
Fs2_control  = 1800;
ws1_control  = 2*(Fs1_control/Fs);
ws2_control  = 2*(Fs2_control/Fs);
BPF_control = fir1(512,[ws1_control,ws2_control]);
PriNoise_control = filter(BPF_control,1,2*randn(N,1));
\end{lstlisting}

\subsection{Loading Primary and Secondary Paths}
The primary path is the path that the noise takes from the noise source to the error microphone, and the secondary path is the path that the anti-noise takes from the secondary source to the error microphone. Both of these paths are essential to the detection of errors. As a result, physical paths are distinct from virtual paths due to the fact that the physical microphones and virtual microphones are situated in distinct locations. In this context, the primary paths and the secondary paths are loaded from files.The length of the primary paths is 128 and the length of the secondary paths is 32.

\begin{lstlisting}[language=Matlab]
% Load physical primary and secondary path.
Pp = load('PrimaryPath_P_4.mat').Pp;
Sp = load('PhysicalPath16.mat').Sp;
% Load virtual primary and secondary path.
Pv = load('PrimaryPath_V_4.mat').Pv;
Sv = load('VirtualPath16.mat').Sv;
\end{lstlisting}

\subsection{Create Disturbance and Filtered Reference Signals}
The following commands generate disturbance signals by filtering the reference signal through the primary path and filtered reference signals by filtering the reference signal through the secondary path. Specifically, different disturbance and filtered reference signals are generated with primary noises in the tuning and control stages, respectively.

\begin{lstlisting}[language=Matlab]
% Creating disturbances and filtered reference signals for tuning noise.
[Dv_tuning,Dp_tuning,Fx_v_tuning,Fx_p_tuning] = CreatReferenceSignal(Pv,Pp,Sv,Sp,PriNoise_tuning,N,L,K,M,J);
% Creating disturbances and filtered reference signals for control noise.
[Dv_control,Dp_control,Fx_v_control,Fx_p_control] = CreatReferenceSignal(Pv,Pp,Sv,Sp,PriNoise_control,N,L,K,M,J);
\end{lstlisting}

\subsection{Tuning Stage 1: Training the Optimal Control Filters}
The following commands train the optimal control filters using the FxLMS algorithm with a stepsize of 0.000001.
\begin{lstlisting}[language=Matlab]
u1 = 0.000001; % Stepsize of the FxLMS algorithm.
[W,Er] = MultichannelFxLMS(L,K,M,N,Fx_v_tuning,Dv_tuning,u1);
\end{lstlisting}

\subsection{Tuning Stage 2: Training the Auxiliary Filters}
The following commands train the auxiliary filters using the LMS algorithm with a stepsize of 0.001.

\begin{lstlisting}[language=Matlab]
u2 = 0.001; % Stepsize of the LMS algorithm.
[H,Er] = AuxiliaryLMS(L,K,J,N,Fx_p_tuning,Dp_tuning,PriNoise_tuning,W,u2);
\end{lstlisting}

\subsection{Control Stage: Training the New Control Filters}
The following commands train the new control filters using the FxLMS algorithm with a stepsize of 0.00001.

\begin{lstlisting}[language=Matlab]
u3 = 0.00001; % Stepsize of the FxLMS algorithm.
[WC,Er,Ev] = ContrFxLMS(L,K,M,J,N,Fx_p_control,Fx_v_control,Dp_control,Dv_control,PriNoise_control,H,u3);
\end{lstlisting}

\subsection{Drawing the Figure of Error Signals}
Drawing the error signals of the four virtual microphones.
\begin{lstlisting}[language=Matlab]
set(groot,'defaultAxesTickLabelInterpreter','latex');
figure  
plot(10*log10(smooth(Ev(1,:).^2,2048,'moving')))
hold on
plot(10*log10(smooth(Ev(2,:).^2,2048,'moving')))
hold on 
plot(10*log10(smooth(Ev(3,:).^2,2048,'moving')))
hold on
plot(10*log10(smooth(Ev(4,:).^2,2048,'moving')))
xlim([0 800001])
ylim([-45 0])
xlabel('Iterations','Interpreter','latex')
ylabel('Square Error (dB)','Interpreter','latex')
title('Time history of error signal at virtual microphones','Interpreter','latex')
legend('Error signal at virtual microphone 1','Error signal at virtual microphone 2','Error signal at virtual microphone 3','Error signal at virtual microphone 4','Interpreter','latex')
grid on
\end{lstlisting}

\begin{figure}[H]
\centering
\includegraphics[width=.7\textwidth]{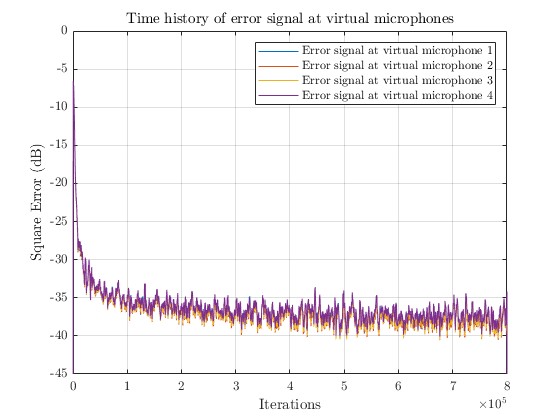}
\caption{ {\small Simulated error signals at four virtual microphones of the MVANC system.}}
\label{figure 5}
\end{figure}
It can be seen from Fig.~\ref{figure 5} that the MVANC system can achieve noise control at the virtual locations and the noise reduction performance is around 40 dB. 

\newpage

\subsection{Drawing the Figure of Control Filter Coefficients}
Drawing the control filter 1 coefficients in both tuning and control stages. The remaining three control filters are very similar to control filter 1.
\begin{lstlisting}[language=Matlab]
figure 
subplot(1,2,1)
plot(1:L,W(1:L),1:L,WC(1:L))
grid on
xlabel('Taps','Interpreter','latex')
ylabel('${{\bf{w}}_{11}}$','Interpreter','latex')
legend('Control filter 1 in the tuning stage','Control filter 1 in the control stage','Interpreter','latex')
xlim([0,512])
subplot(1,2,2)
[H, F] = freqz(W(1:L), 1, 1024, Fs);
plot(F, abs(H));
hold on
[H, F] = freqz(WC(1:L), 1, 1024, Fs);
plot(F, abs(H));
grid on
xlim([0,3000])
xlabel('Frequency (Hz)','Interpreter','latex')
ylabel('$|{W_{11}}(f)|$','Interpreter','latex')
\end{lstlisting}

\begin{figure}[H]
\centering
\includegraphics[width=1\textwidth]{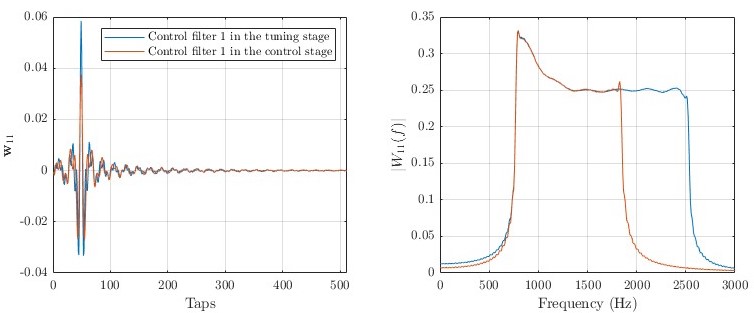}
\caption{ {\small Simulated control filter 1 coefficients of the MVANC system.}}
\label{figure 6}
\end{figure}

It can be seen from Fig.~\ref{figure 6} that the control filter 1 derived from the tuning stage exhibits a passband within the 800 to 2500 Hz
frequency range. Conversely, control filter 1 from the control stage demonstrates a passband within the narrower 800 to 1800 Hz range. This result is consistent with the different primary noises used in the tuning and control stages.

\newpage
\section{Conclusion}
The purpose of this article is to provide a comprehensive explanation of a MATLAB program that simulates the feedforward multichannel virtual sensing active noise control (MVANC) technique by employing the multichannel filtered-x least mean square (MCFxLMS) technique. 

\printbibliography


\end{document}